# Design of Near Infrared Sky Brightness Monitor and Test Running at Ngari Observatory in Tibet

Qi-Jie Tang[1], Jian Wang[1,†], Shu-cheng Dong[1], Jin-ting Chen[1], Yi-hao Zhang[1], Feng-xin Jiang[1], Zhi-yue Wang[1], Ya-qi Chen[1], Ming-hao Jia[1], Jie Chen[1], Hong-fei Zhang[1], Qing-feng Zhu[2], Peng Jiang[3,*], Tuo Ji[3], Shao-hua Zhang[3], Yong-qiang Yao[4], Yun-he Zhou[4], Hong-shuai Wang[4], Peng Tang[4]
(1. State Key Laboratory of Technologies of Particle Detection and Electronics, Department of Modern Physics, University of Science and Technology of China, Hefei 230026, China
2. Department of Astronomy, University of Science and Technology of China, Chinese Academy of Sciences, Hefei 230026, China
3. Polar Research Institute of China, Shanghai 200136, China
4. National Astronomical Observatories, Chinese Academy of Sciences, Beijing 100012, China)

†Email: wangjian@ustc.edu.cn
*Email: jiangpeng@pric.org.cn

**ABSTRACT:** Tibet is known as the third pole of the earth, as high as the South Pole and North Pole. The Ngari (Ali) observatory in Tibet has the advantage of plenty of photometric night, low precipitable water vapor, high transmittance, good seeing. It is a good site, and promising to be one of the best place for infrared and submillimeter observations in the world. However, there is no data available for sky background brightness in such place, which restrict the astronomical development of the sites. In the near infrared band of J, H, Ks, a NIR sky brightness monitor (NISBM) is designed based on InGaAs photoelectric diode. By using the method of chopper modulation and digital lock-in amplifier processing, the SNR (Signal Noise Ratio), detectivity and the data acquisition speed of the device is greatly improved. For each band of J, H, Ks, an independent instrument is designed and calibrated in laboratory. The NISBM has been installed in Ngari observatory in July of 2017 and obtained the first data of NIR sky brightness at Ngari observatory.

# 1. Introduction

The sky background radiation is played as important role in the astronomical observations, which can influence the observation of whole magnetic and visible spectrum. Sky background radiation from ultraviolet to far infrared on ground and in space has been studied fully. There are different emission principle in different waveband[1]. In the NIR waveband, the main contribution to the sky background radiation is emission of hydroxyl radical (OH) airglow in the atmosphere, which affect the astronomical observations more than the one in visible band[2][3]. The airglow is emitted by the vibrational de-excitation of excited OH molecules created in the mesosphere at an altitude of 80-105km. The excited OH radical cascades radiatively to lower energy states and produces the characteristic airglow spectrum in NIR band [4]. J, H, Ks are the three narrow windows of NIR band in the atmosphere as shown in Fig.1, in which the main flux is mainly emission from OH airglow and only at the end of NIR band which is about ~2.3um the thermal emission from sky start to dominate.

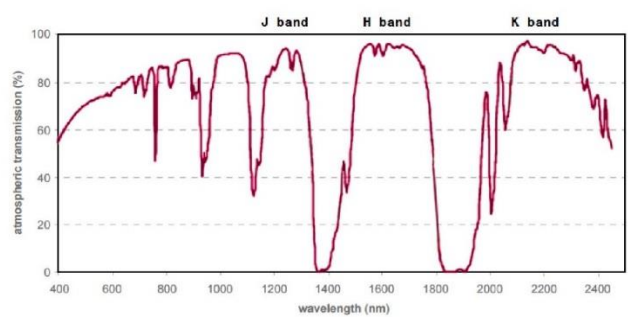

Fig.1.Definition of NIR band

As shown in Fig.1, the most of NIR light is absorbed by the vaper in the atmosphere. Anyway, there are still narrow windows the NIR light can reach to ground.

Recently, development of the infrared astronomy has been stressed in China and several optical and infrared telescopes is proposed in Antarctic and mainland such as 2.5-meter KDUST (Kunlun Dark Universe Survey Telescope) in Antarctic[11], 12-meter optical and infrared telescope in mainland[12]. Tibet is known as the third pole of the earth, as high as the South Pole and North Pole. The Ngari (Ali) observatory in Tibet has the advantage of plenty of photometric night, low precipitable water vapor, high transmittance, good seeing. It is a good site, and promising to be one of the best place for infrared and submillimeter observations in the world[9]. Furthermore, the Ngari observatory also can be the base of test observing for instruments and telescopes installed in Antarctic.

Only at the site where the background infrared radiation is weak, an observatory can be built. Hence, the background infrared radiation measurement should be carried out at the candidate site. However, all candidate sites for future Chinese astronomical telescopes with infrared capabilities, including sites at Antarctic, Ngari observatory in Tibet province and Daocheng Observatory in Sichuan province, such measurements are not complete. A series of important features of a telescope can be crucially restrained by the flux of the sky brightness, including the survey depth, the exposure time of astronomical imaging system. Near infrared is an important wavelength range for astronomy. Knowledge about infrared observing conditions, especially the average intensities and variations could be referenced for further design of infrared telescope and its instrumentation. Meanwhile, the development of infrared sky brightness measurement system has a crucial effect for the design of infrared observing project.

So far, there are some significant works on the sky brightness measurement including measurement of the OH sky radiation and improvement of measurement sensitivity of NIR spectra[5-8]. In these work, cryocoolers or solid Nitrogen are used to cryogenically cool the detector to lower the noise. Filter wheel is used to switch the filter in the cryogenic environment. To develop these instruments, many complex technologies including vacuum technology, cryogenic technology and cryogenic electronics are used. They used the InSb detector which should work in the cryogenic environment such as 77k. We studied the most common detector for infrared detection such InSb, InGaAs and HgCdTe and found the InGaAs detector is more suitable for NIR detection in J, H, Ks waveband.

Based on the InGaAs detector, oriented to the harsh environment of Nagri Observatory in Tibet and Kunlun station in Antarctica, the NIR sky brightness monitor (NISBM) is designed, including optics, low noise readout electronics, mechanics, electronical control and automatic observation. The NISBM was installed successfully and get the first light at the end of July, 2017. In Section 2, we describe the whole system structure with the detector, NIR filters, optics, and mechanics. To obtain amplitude of the sky brightness, a weak signal readout system is introduced in section 3. In the Section 4, the calibration is described and finally, the measurement result of Nagri Observatory is presented in Section 5.

## 2. System Structure

Since the NISBM works in JHKs waveband, the filters of JHKs band and the detector should be chosen properly. According to the definition of JHKs board band, we selected the filters as shown in Table1.

Table1 JHKs board filters

| band | center | cut-on | cut-off |
|------|--------|--------|---------|
| J | 1.250μm | 1.170μm | 1.330μm |
| H | 1.635μm | 1.490μm | 1.780μm |
| Ks | 2.150μm | 1.990μm | 2.310μm |

The transmittance of the filters as shown in Fig.2.

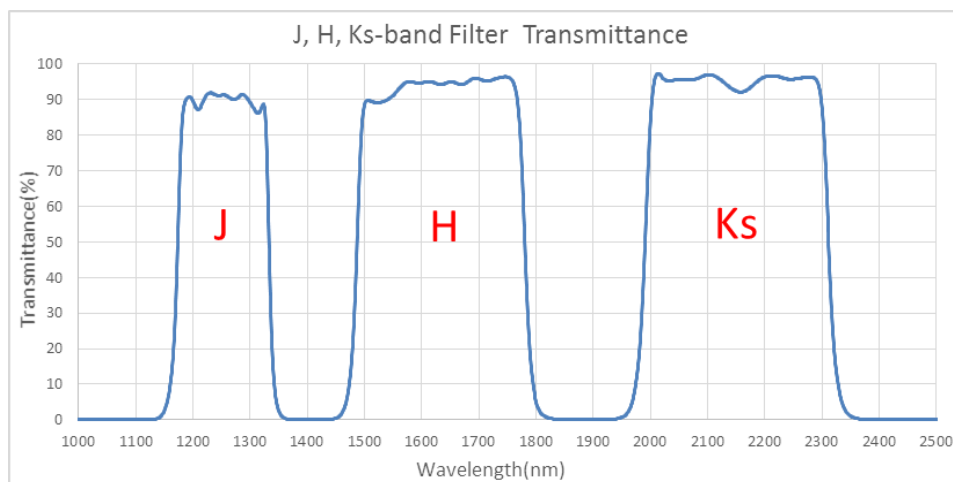

Fig.2. Transmittance of the JHKs filters

Comparing the main NIR detectors including the InGaAs detector, InSb detector, HgCdTe detector as shown in Table 2, the InGaAs detector is chosen for its higher detectivity (D*) and lower noise in JHKs waveband. Furthermore, it use TEC cooling to cool detector to about -40℃, which makes our instrument design simple. If we use InSb detector or HgCdTe detector, a cryocooler should be used to cool detector, which will make the instrument design more complex.

Table 2 Comparison of 3 main detectors for NIR detection

| Detector | InGaAs | InSb | HgCdTe |
| --- | --- | --- | --- |
| wavelength range(μm) | 0.9~2.55 | 1.0~5.5 | 2~5 |
| working temperature | -40℃~room temperature | 77K | 77K |
| $D^*(cm \cdot Hz^{1/2}/W)$Typical | $4.5 \cdot 10^{11}$ | $1 \cdot 10^{11}$ | $8 \cdot 10^{10}$ |

For InGaAs detectors, there are different models with different cut-off wavelengths as shown in Fig.3 in which the spectral response was measured in room temperature. After noise test of InGaAs detector in Lab, we found that the noise is enough low to obtain the weak sky brightness when the detector is cooled to -40℃ and don't need to be cryogenically cooled such as 77K.

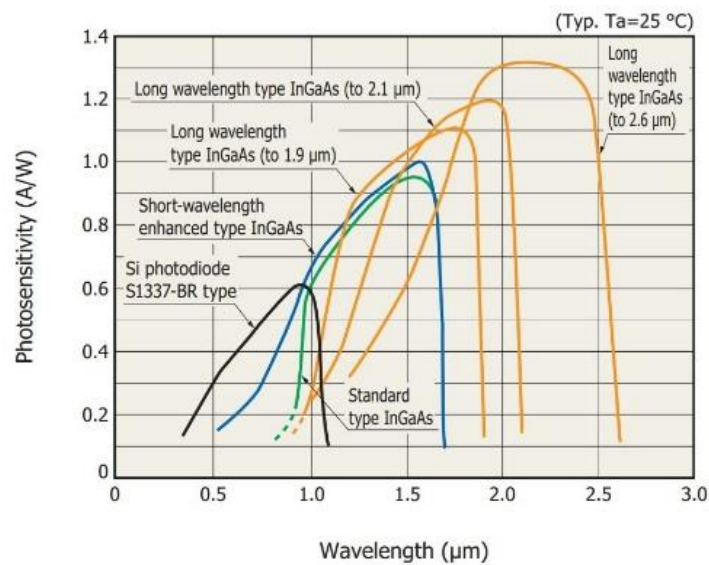

Fig.3. Spectral response with different type InGaAs detectors

To improve the efficiency of measurement, a stand-alone instrument with its own optics, mechanics and detector is designed for every filter such as J, H, Ks. It will be more convenient if the NISBM is installed in Antarctica with less moving component such as filter wheel, which means less possibility of instrument failure. The design of optics is same for the instruments with 3 different filters as shown in Fig.4 which consists of a scanning mirror, an optical window, an aperture stop, an aspheric lens, an NIR filter and an InGaAs detector.

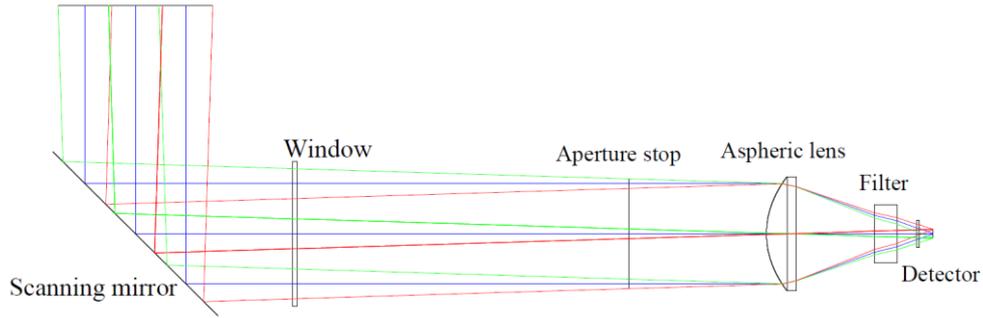

Fig.4. Schematic design of optics

In our design, different filters are used in the different waveband as shown in Fig.2. For the weakness of sky brightness, to improve the performance of measurements, the type of InGaAS detectors with the best detectivity in corresponding waveband is chosen to get the best SNR (Signal Noise Ratio). As shown in Table 3, the longer the cutoff wavelength the larger the readout noise and the bigger the dark current. So according to the range of wavelength of JHKs, the model of detector with the smallest cutoff wavelength is choose. The model of InGaAs detector with 1.7um cutoff wavelength is choose for measurement of J band, the model with 2.05um cutoff wavelength is for H band and the model with 2.55um cutoff wavelength is for Ks band. Each has the best photosensitivity in its waveband of measurement. For each of the three instruments of sky brightness measurement in JHKs waveband, the structure is same but the filters and position of the detectors. The position of the detectors is adjusted slightly according to the different waveband.

Table 3. The parameters of InGaAs detectors

| Model | G121180 | G121182 | G121183 |
|---|---|---|---|
| Work wavelength range($\mu$m) | 0.9~1.7 | 0.9~2.05 | 0.9~2.55 |
| Waveband of measurement($\mu$m) | J(1.17~1.33) | H(1.49~1.78) | Ks(1.99~2.31) |
| Dark Current(nA) | 1.5 | 25 | 2100 |
| Detectivity-D*(cm·Hz$^{1/2}$/W) | $6.3 \cdot 10^{12}$ | $2 \cdot 10^{12}$ | $4.5 \cdot 10^{11}$ |
| Noize quivalent power ENP(W/Hz$^{1/2}$,peak) | $7.5 \cdot 10^{-14}$ | $9 \cdot 10^{-14}$ | $4 \cdot 10^{-13}$ |

The optics of instruments are simulated by ZEMAX in different waveband. The simulations of spot diagram are shown in Fig. 5. The RMS values of spot diameter of three waveband are less than 50um in the whole FOV (field-of-view), which are much less than the effective detective diameter of the detector. The footprint diagrams of FOV in three waveband are shown in Fig.6. The light from the whole FOV can be focused within the detective range of detector with its diameter of 2mm. The encircled energy are shown in Fig.7 and 90% energy is focused on the 70um range. The simulation results show the optics meets the demand that the effectiveness of gathering energy is optimal in the FOV of 3.5°.

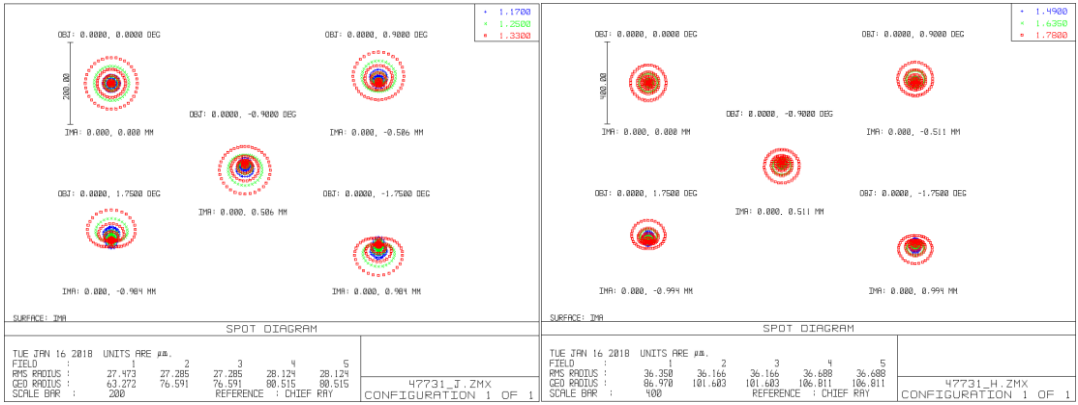
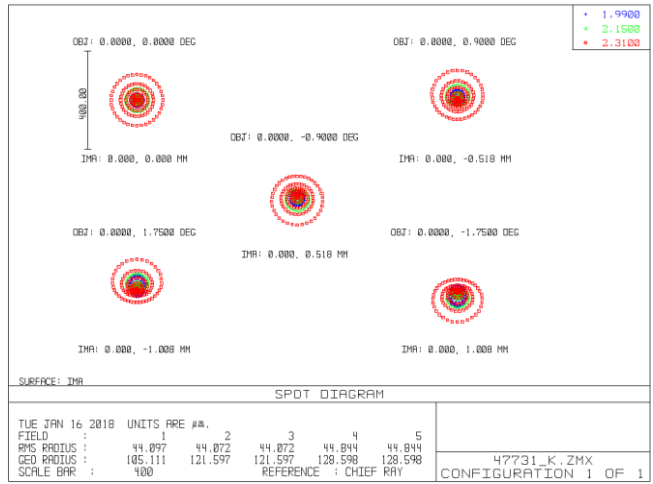

Fig.5. Spot diagram of three instruments in J,H,Ks band

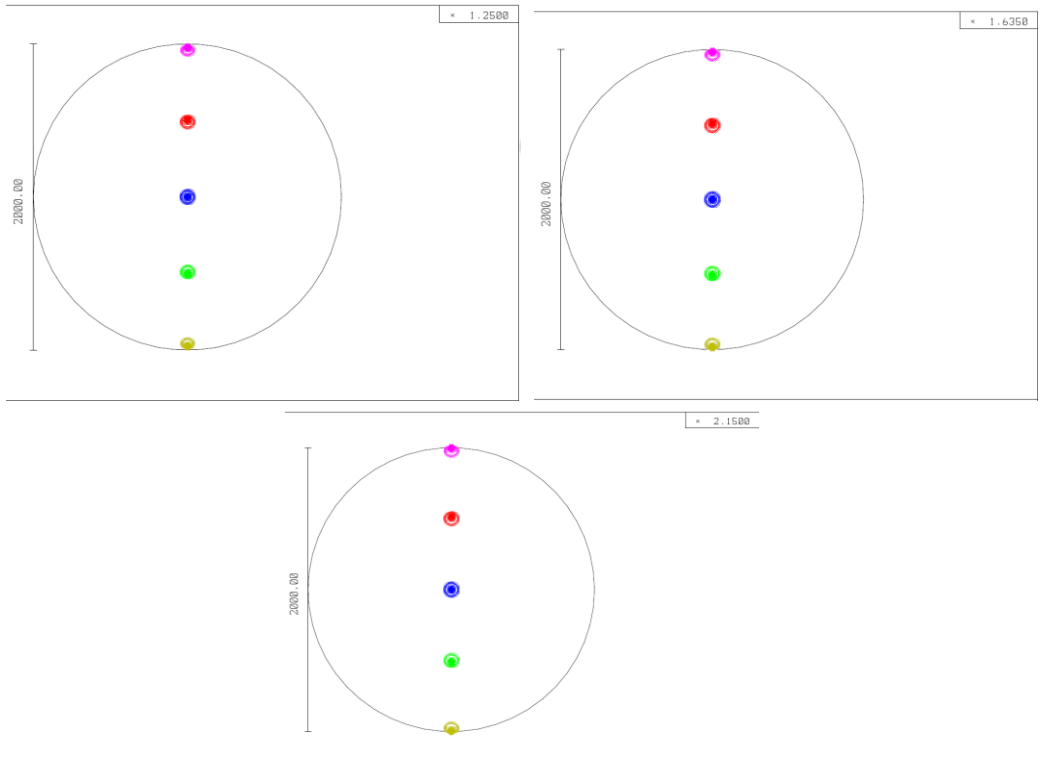

Fig.6. Footprint diagram of three instruments in J,H,Ks band

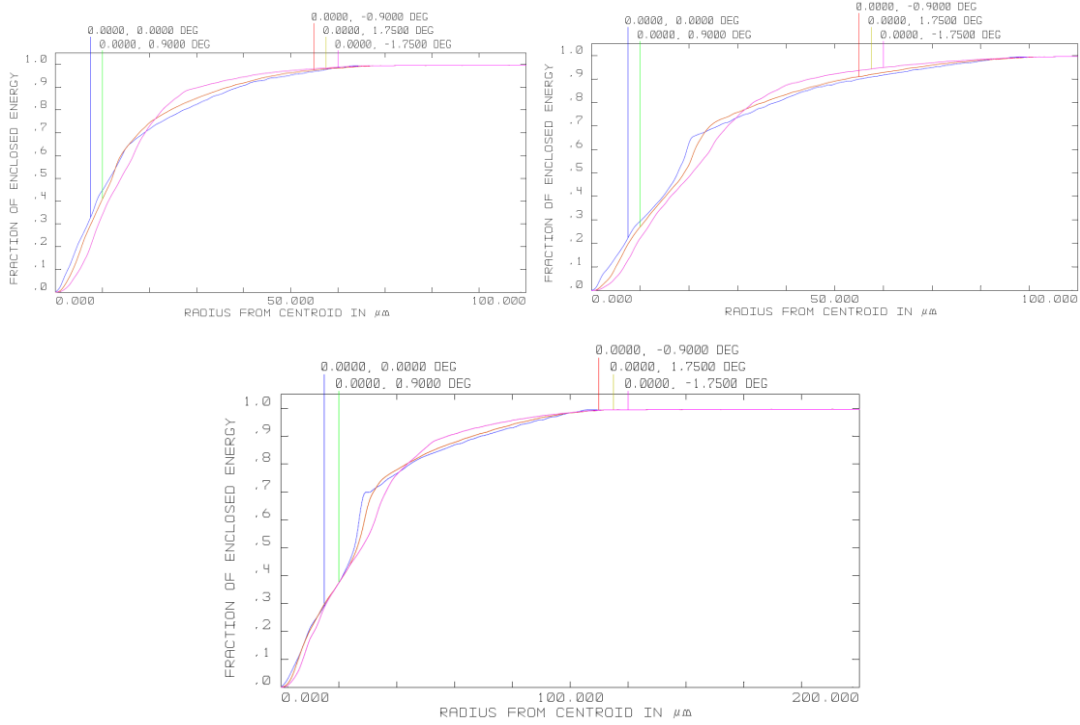

Fig.7. Encircled energy of three instruments in J,H,Ks band

The parameters of three instruments are shown in Table 4.

Table 4. The parameters of JHKs measurements

| Parameter | J band | H band | Ks band |
| --- | --- | --- | --- |
| Aperture Diameter | 22.0mm | 22.0mm | 22.0mm |
| f/# | 1.45 | 1.45 | 1.45 |
| Field-of-view | 3.5 degrees | 3.5 degrees | 3.5 degrees |
| Sky Area (Solid Angle) | $1.25 \cdot 10^8 \text{arcsec}^2$ | $1.25 \cdot 10^8 \text{arcsec}^2$ | $1.25 \cdot 10^8 \text{arcsec}^2$ |
| Filter Center Wavelength | 1.250μm | 1.635μm | 2.150μm |
| Filter Bandwidth (Full-Width-Half-Max) | 0.160μm | 0.290μm | 0.320μm |
| Filter Throughput | ~0.90 | ~0.94 | ~0.95 |
| Average Responsivity | ~0.84 A/W | ~1.12 A/W | ~1.27 A/W |
| Flow Response(measured)(μJy $\text{arcsec}^{-2}$/pA) | 86.1 | 70.0 | 99.3 |

According to the optics design, the mechanics and its control are designed. For the signal of NIR sky brightness is slow-changing and almost constant, to improve the SNR, a chopper is used to modulate the signal to AC. The structure of each instrument is shown in Fig.8.

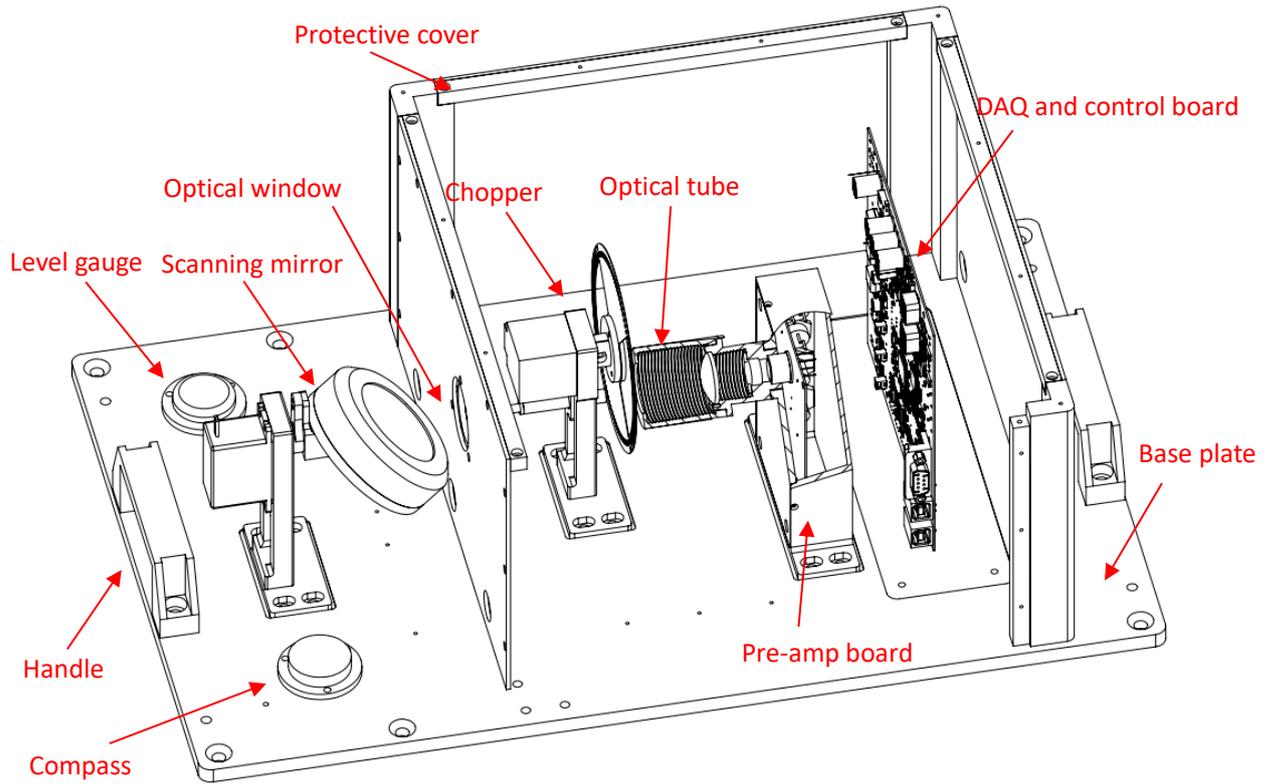

Fig.8. The whole structure of each instrument

The mechanics include an optical tube, a protective cover, a base plate and the fixed structure of the detector, the chopper and the scanning mirror. The control components include the motor driver circuit of the chopper and its control, the motor driver circuit of the scanning mirror and its control. The scanning mirror can be directed from 0 to 180 degrees along the sky, which can be controlled by operational software on the computer. The same as the chopper.

## 3. Electronics and its DAQ

The whole electronics are shown in Fig.9.

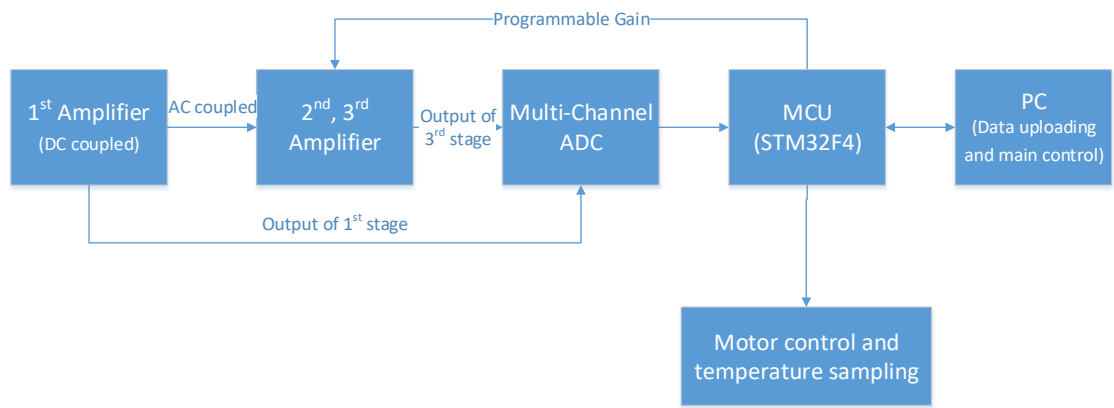

Fig.9mei. Electronics design of each instrument

The output of the detector is weak current including the DC base component and AC modulated component. The amplifier has 3 stages as shown in Fig.10. The 1st sage is a DC-coupled TIA (trans-impedance amplifier) which convert the weak current to voltage with high-gain. The gain of 1st stage is very large and the noise performance of whole electronics is depended on the 1st stage. The 2nd stage is an AC-coupled PGA (programmable gain amplifier) which can block the DC component and amplify the AC component with programmable gain. The 3rd stage is a 2-order active LPF (low pass filter) which can filter the high-frequency noise and limit the bandwidth of signal.

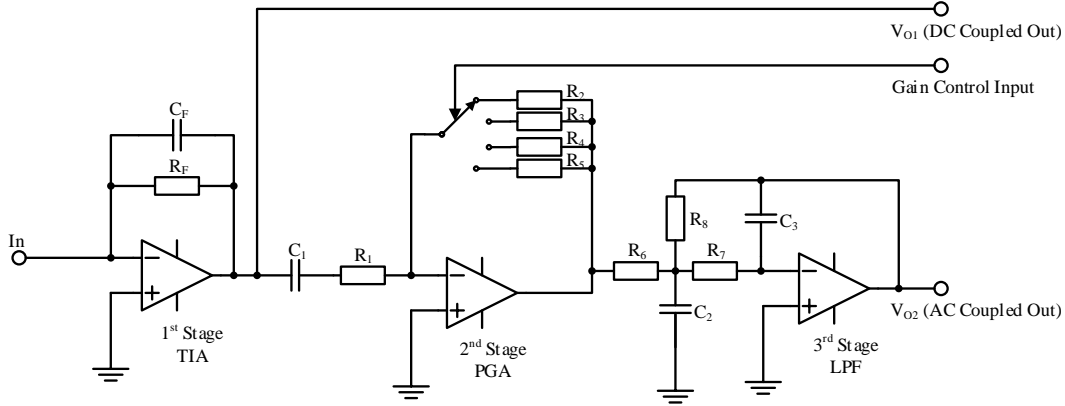

Fig. 10. The schematic diagram of the high-gain amplifier

The output of 1st stage is

$$V_{O1} = I_{in} \times R_F \quad (1)$$

The noise performance (SNR) of the amplifier is dependent on the 1st stage. The SNR of 1st stage is proportioned to $\sqrt{R_F}$, which means the bigger the gain ($R_F$) is, the better the SNR is. The gain is limited by the strength of DC base component which comes from the radiation of the instrument. The power rail of the amplifier is ±5V and the gain should be increased as high as possible under the unsaturated condition. In the 2nd stage, the DC component is blocked and the AC component is amplified with 8 level programmable gain. The 3rd stage is a 2-order active LPF whose bandwidth is configured according to the frequency of the chopper, which can filter the high-frequency noise.

After amplification and filtering, the signal is put into the ADC (Analog-to-Digital Converter) and digitalized. It is processed by digital phase lock algorithm after reducing the 1/f noise and DC drifting of operational amplifier as shown in Fig.11. The digital LIA (Lock-in amplifier) consists of HGA (high-gain amplifier), ADC and digital phase lock module in MCU (Microcontroller Unit) and are called Orthogonal vector LIA [10]. The digital phase lock algorithm is implemented in MCU including PSD (Phase-Sensitive Detector) with the reference clock, LPF, calculation of the amplitude and phase.

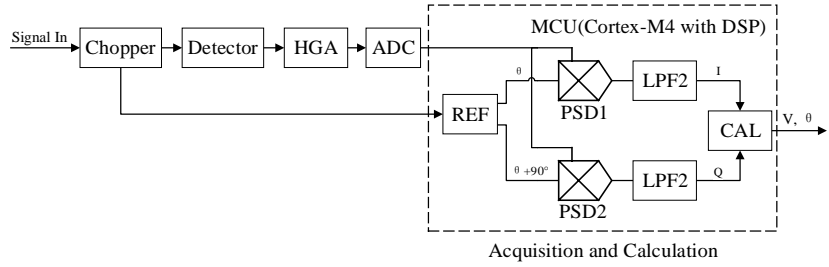

Fig. 11. Structure of digital LIA. (HGA: High-gain amplifier; ADC: Analog to Digital converter; PSD: Phase-Sensitive Detector; LPF: Low-Pass Filter; REF: Reference Clock Generator; CAL: Calculation)

Orthogonal vector LIA can output in-phase and orthogonal components simultaneously. It needs two PSD which the signal input are the same but the reference input on the phase difference of 90 °. PSD1 reference input in the in-phase channel has the phase shift of $\theta$ while orthogonal channel PSD2 reference input has $\theta$ +90 ° phase shift. The in-phase output of the orthogonal vector LIA is:

$$I = V\cos\theta \quad (2)$$

And its orthogonal output is:

$$Q = V\sin\theta \quad (3)$$

The amplitude V and phase $\theta$ of the measured signals can be calculated by the two outputs:

$$V = \sqrt{I^2 + Q^2} \quad (4)$$

$$\theta = \arctan(Q/I) \quad (5)$$

Orthogonal vector LIA use two orthogonal components to calculate the amplitude V and phase θ, which can avoid the phase shifting of the reference clock and accuracy of measurement from phase shifting.

According the chopper design and its size, the signal waveform after chopping is calculated as shown in Fig.12, which has flat form on top/bottom side. The difference value of two flat forms is the value that the signal of sky brightness subtracts the radiation signal of chopper vanes. The signal shown in Fig.14 has harmonic components while we use LIA method to remove the harmonic components and get the fundamental component only which the amplitude is 1.18 calculated by Fourier transform.

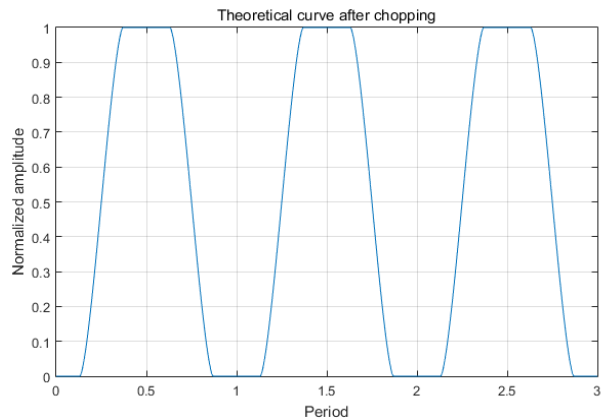

Fig.12. Calculated output waveform after chopping

The digital algorithm simplifies the circuit of the whole electronics and can adjust the time constant of the LPF conveniently to realize different equivalent bandwidths. If the intensity of signal is high enough, the response speed of the measurement can be improved by reducing the time constant. When the signal is weak, the bandwidth can be reduced and the SNR is improved by increasing the time constant. The algorithm has the advantages of high accuracy, flexible application, simple implementation, low computation and low memory capacity.

## 4. Calibration

According the Planck formula:

$$I(\lambda, T) = \frac{2hc^2}{\lambda^5} \cdot \frac{1}{e^{\frac{hc}{\lambda kT}} - 1} \quad (6)$$

When the radiation from extended area blackbody with temperature T covers the FOV, the current output $i(T)$ from detector is calculated as:

$$i(T) = AS\overline{R}C_{Chopper}\eta_{Optics} \int_{\lambda_1}^{\lambda_2} I(\lambda, T)\eta(\lambda)d\lambda \quad (7)$$

According the discussion of NIR sky brightness measurement above, the parameters is obtained as followed:

(1) A is solid angle. The field angle $\theta$ is 3.5°.

$$A = 2\pi\left(1 - \cos\frac{\theta}{2}\right) = 2.9305 \cdot 10^3 rad^2 \quad (8)$$

(2) S is the effective entrance pupil area of the optical window. The diameter D of the effective entrance pupil is 0.022m.

$$S = \pi(\frac{D}{2})^2 = 3.8013 \cdot 10^{-4} m^2 \quad (9)$$

(3) $\overline{R}$ is the average responsivity of the detector which is about 1A/W.
(4) $C_{Chopper}$ is the ratio of sinusoidal component and peak value of signal, which is about 1.18 after calculation which is discussed in the end of section 3.
(5) $\eta_{Optics}$ is the total transmission coefficient of the lenses, the optical window and the scanning mirror. $\eta_{Optics} \approx 0.98 \cdot 0.85 \cdot 0.9 = 0.75$.
(6) $\eta(\lambda)$ is the transmission of the filters which from the manufacturer as shown in Fig.2.
(7) $\lambda_1$ and $\lambda_2$ are the boundary of bandwidth.

Then we can calculate the current value of output from detector vs temperature of blackbody theoretically in the JHKs band according to the formula (1). The curves are shown in Fig.13.

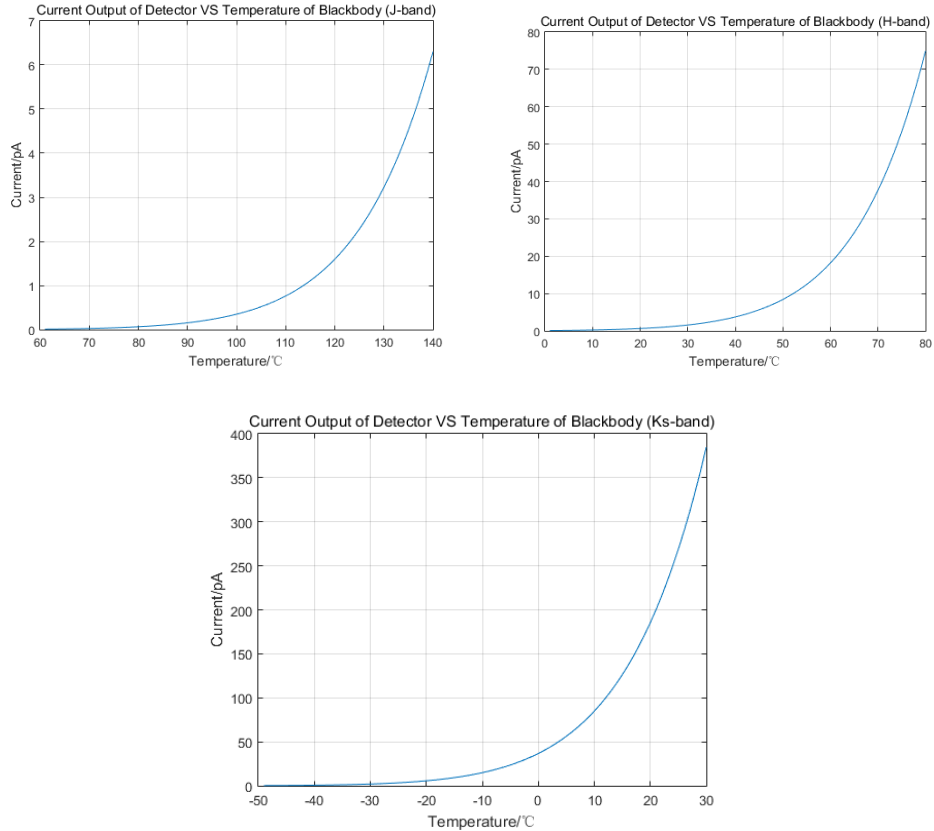

Fig.13. The curves of current output from detector vs temperature of blackbody in J,H,Ks band

Seen from Fig.13, when the temperature is less than 70 degrees Celsius, the signal from detector in J band is less than 0.02pA which can be neglected. In the H band the signal is less than 0.2pA when the temperature is less than 10 degrees Celsius, which means the radiation of ambient has some influence of the measurement. When the temperature is lower, the influence is smaller. In the Ngari Observatory, the temperature of summer night is about 5~10 degrees Celsius. The radiation of background in the Ks band will very strong which can be 30~64pA. The radiation intensity from the vanes of the chopper is similar with the one from sky, even stronger. Since there is no blackbody to calibrate the measurement in Ks band, we still can calibrate the measurement through temperature measurement of chopper vanes.

The real signal from sky brightness is $i_{Sky}$ which can be calculated by

$$i_{Sky} = i_{Chopper} \pm i_{Signal} \quad (10)$$

In the formula (10), the $i_{Chopper}$ is the signal from chopper vanes and $i_{Signal}$ is the signal we have measured. The sign of $\pm$ is decided by phase of $i_{Signal}$. The sign + represents that signal strength of sky brightness is stronger than that of chopper vanes. The sign – represents that signal strength of sky brightness is weaker than that of chopper vanes.

The chopper is not the ideal blackbody and its emittance is less than 1. If we want to get the accurate relation between radiation of chopper vanes and its temperature, blackbody calibration experiments should be conducted.

Furthermore, the parameters of instrument have their deviation from real value. The real current value is proportional to the theoretical value, the formula of which is:

$$S(T) = ai(T) + b \quad (11)$$

In the formula (11), S(T) is the real current value from detector; a is the deviation coefficient of system responsivity and its theoretical value, which is decided by the instrument and not influenced by ambient condition; b is the offset, which is the value of $i_{Chopper}$ from chopper vanes and influenced by temperature of ambient. a and b is the calibrated coefficients which is obtained by blackbody calibration experiments and introduced below.

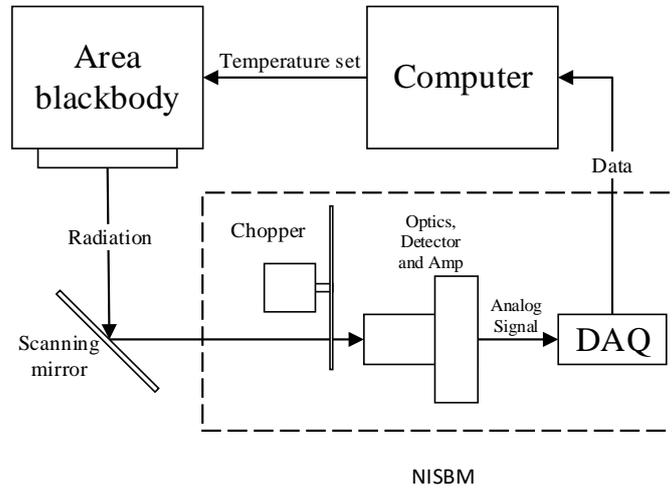

Fig.14. The setup of blackbody calibration experiments. NISBM: NIR sky brightness monitor; DAQ: Data Acquisition board

The setup of blackbody calibration is shown in Fig. 14. An extended area blackbody is used as the source and covers the FOV of the instrument. Adjusting the temperature of blackbody, the current of detector is measured and recorded after the blackbody has its steady temperature. According to the parameters of three instrument in J,H,Ks band, the theoretical current value is calculated from formula (7). According to the formula (11), the coefficient a and b can be fitted as shown in Fig.15.

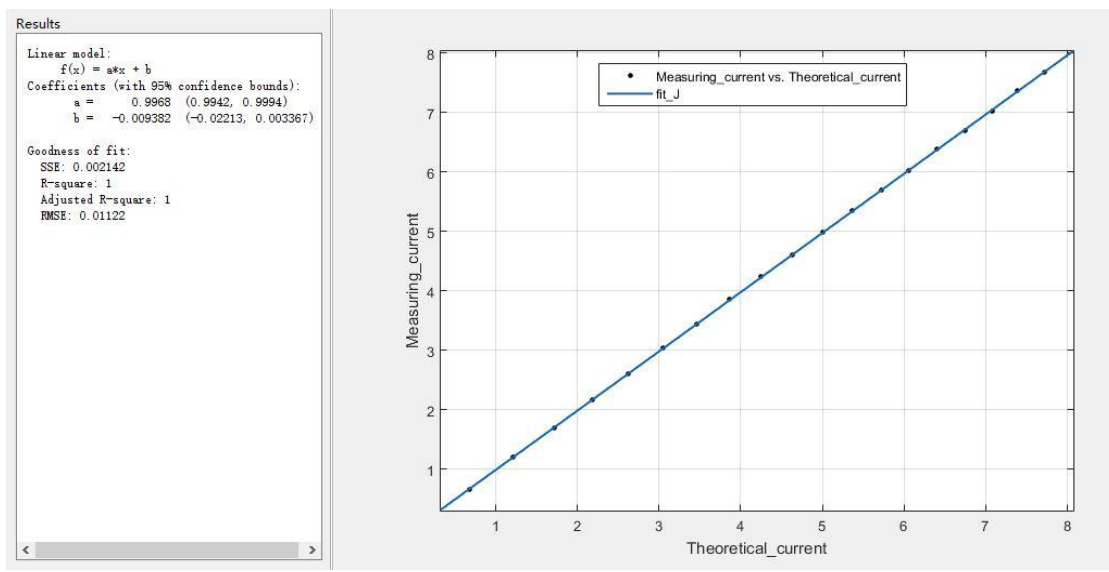

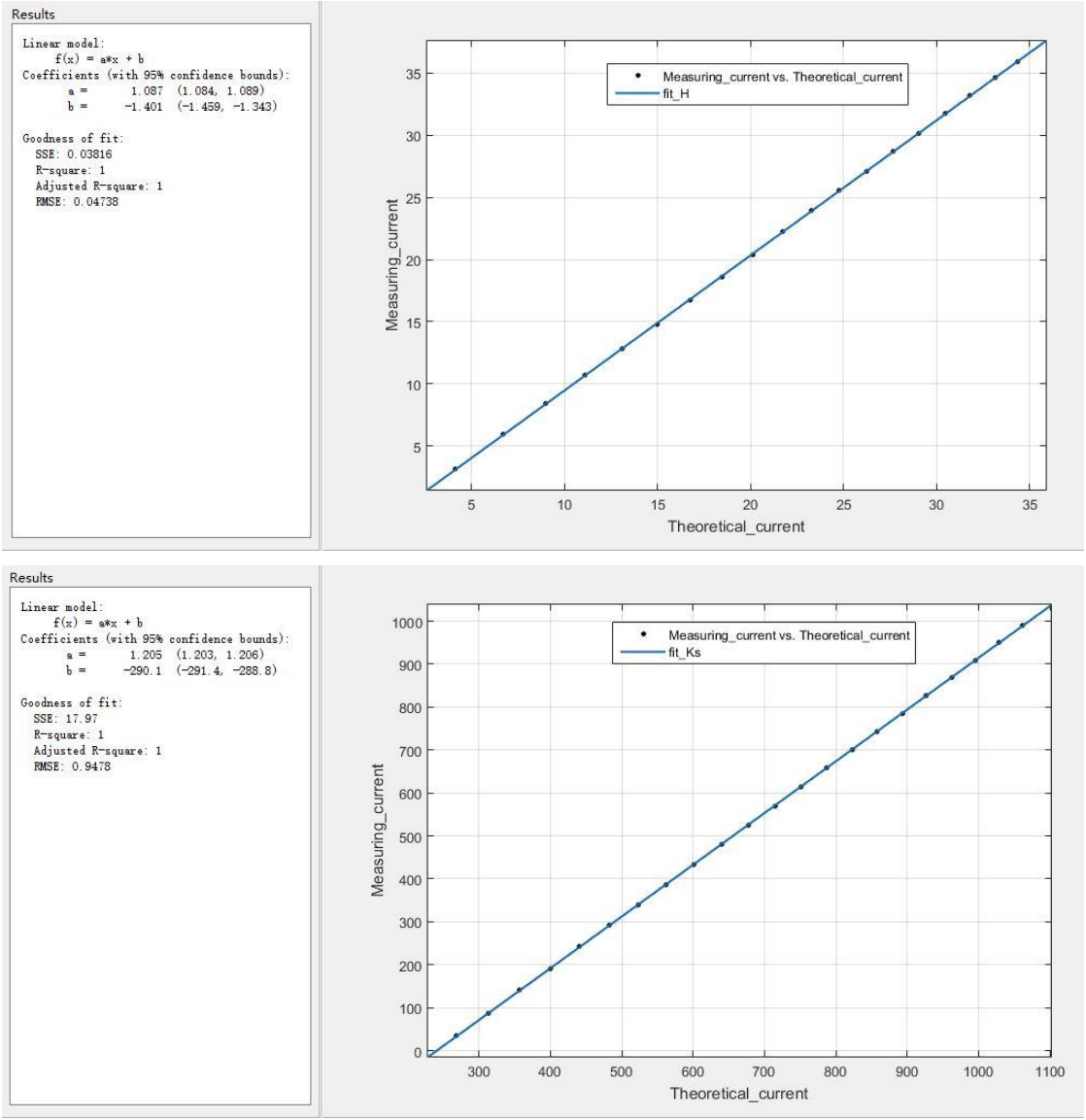

Fig.15. Fitting curve of blackbody calibration in J,H,Ks band

The results of fitting are shown in Table 5.

Table 5. Fitting coefficient in J,H,Ks band

| Coefficient | J band | H band | Ks band |
|---|---|---|---|
| a | 0.9968 | 1.087 | 1.205 |
| b | -0.0094 | -1.401 | -290.1 |
| R-square | 1 | 1 | 1 |
| RMSE (root mean square error) | 0.01122 | 0.04738 | 0.9478 |

Seen from Fig.15 the linearity of fitting is very good. The coefficient a is about 1 which means the theoretical calculation and value from experiment match well. The temperature of ambient in experiments is about 27℃. Seen from the table 5, the coefficient b in J band is almost 0, which means in the room temperature the radiation from background in J band is very small. The coefficient b in H band is 1.401pA, which means in the room temperature the radiation from background in H band is still small. When the temperature in ambient decreases, the value also decreases. The coefficient b in Ks band is 290.1pA, which means the radiation strength of background is strong and consistent with the theoretical analysis.

If the average flux density from sky is L(f), the current output from detector $i_{Signal}$ calculated by

$$i_{Signal} = aAS\overline{R}C_{Chopper}\eta_{Optics}\overline{\eta}L(f)\Delta f + b \quad (12)$$

In the formula (12), $\overline{\eta}$ is the average transmission of the optics, $\Delta f = f_2 - f_1$ is the frequency range of the filter.

The unit of flux is $\mu Jy\ arcsec^{-2}$.

$$1\mu Jy\ arcsec^{-2} = 10^{-6} \cdot 10^{-26} W/(m^2 \cdot Hz \cdot arcsec^2)$$
$$= 1 W/(m^2 \cdot Hz \cdot rad^2) \cdot [10^{-34} \cdot \left(\frac{180 \cdot 3600}{\pi}\right)^2] \quad (13)$$

From the formula (12) and (13), L(f) is calculated by

$$L(f) = \frac{1}{aAS\overline{R}C_{Chopper}\eta_{Optics}\overline{\eta}\Delta f \cdot [10^{-34} \cdot \left(\frac{180 \cdot 3600}{\pi}\right)^2]} \cdot (i_{signal} - b)$$
$$= C(i_{signal} - b) \quad (14)$$

In the formula (14), C is the calibrated coefficient which is decided by instruments. The results of calibration experiments in J,H,Ks band are shown in Table 6.

Table 6. The results of calibration experiments in J,H,Ks band

| Parameter | J band | H band | Ks band |
|---|---|---|---|
| a | 0.9968 | 1.087 | 1.205 |
| $\overline{\eta}$ | 0.90 | 0.94 | 0.95 |
| $\Delta f(Hz)$ | $3.087 \cdot 10^{13}$ | $3.337 \cdot 10^{13}$ | $2.098 \cdot 10^{13}$ |
| $C(\mu Jy\ arcsec^{-2}/pA)$ | 86.1 | 70.0 | 99.3 |

When the temperature of ambient is higher, the radiation in Ks band will be stronger. The radiation from chopper vanes may be greater than that of sky brightness. Furthermore, the variation of temperature of chopper vanes will have a big influence on measurement. So the temperature of chopper vanes should be measured to fix the influence, which means the relation between coefficient b and temperature should be obtained.

The chopper is driven by stepper motor. The thermal generated by stepper motor when it works will be conducted to chopper vanes. The temperature of chopper vanes will higher than that of ambient. Because of sealing instrument and thermal balance inside after steady working, the temperature difference from different part will be almost constant. It is difficult to measure the temperature of chopper vanes when it is working. So we can get the temperature of chopper vanes by measuring the temperature of ambient inside instrument indirectly. From the experiments, the temperature of chopper vanes is 2 degrees Celsius higher than that of ambient near the chopper vanes. That means the temperature of ambient near the chopper vanes is measured, the temperature of chopper vanes is obtained.

The blackbody calibration experiments in Ks band were conducted in the ambient with different temperatures. The coefficient a and b were fitted as shown in Table 7.

Table 7. The corrected value of a and b of Ks band in different temperature

| Ambient temperature (°C) | temperature of Chopper vanes (°C) | a | b |
|---|---|---|---|
| 27.0 | 29.0 | 1.205 | 290.1 |
| 8.5 | 10.5 | 1.204 | 73.15 |
| 8.0 | 10.0 | 1.204 | 70.36 |

The results of experiments were fitted with curve of blackbody response, we found the coefficient b is equal to the radiation of blackbody in the same temperature. In the experiment, the optical path of radiation of blackbody has one component more than that of chopper vanes, the component is the optical window. That means the emittance of chopper vanes is same as the transmission of optical window which is about 0.85. According to the formula (7), the b (T) will be calculated with $\eta_{Optics}$ which is the total transmission coefficient of the lenses and the scanning mirror without the optical window.

Some deviation will be got in this method, but it is very limited as shown in Table8. In the winter night in Nagri Observatory, the range of temperature is -5~-35℃. In our method the temperature of chopper vanes has deviation. If the maximum deviation of temperature is 0.5℃, the calculated error can be obtained as shown in Table 8 and the flux of sky brightness is 1000μJy $arcsec^{-2}$

Table 8. The calculated error of different temperature in Ks band

| Ambient temperature/℃ | Deviation of correction/μJy $arcsec^{-2}$ | Deviation of correction /% |
|---|---|---|
| 0 | 131.5 | 13.1 |
| -10 | 57.6 | 5.8 |
| -20 | 23.6 | 2.4 |
| -30 | 9.0 | 0.9 |
| -40 | 3.1 | 0.3 |

## 5. Test running at Nagri Observatory

In July, 2017, the three NIR sky brightness monitors is installed successfully at Nagri Observatory, Tibet as shown in Fig. 16. and the data of first light was obtained.

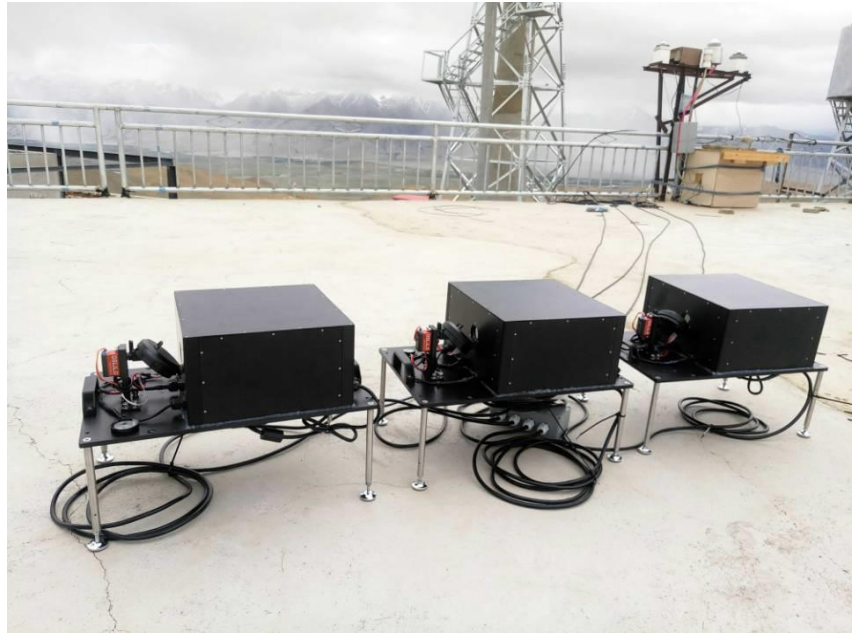

Fig. 16. NISBM installed at Nagri Observatory

In the night of Oct. 16, the weather in Nagri is sunny, almost no cloud. The temperature is 8.2~5.2℃. The data of Ks was corrected using the method discussed in Section 4. The temperature and corrected value is shown in Fig.17.

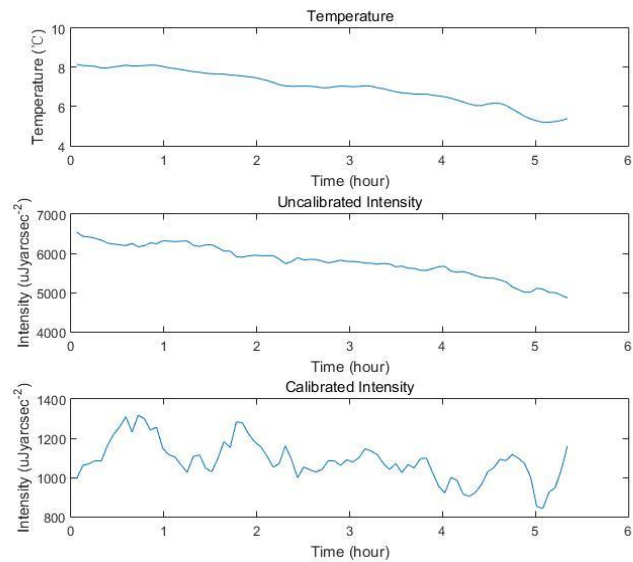

Fig.17. The temperature and the corrected value in Ks band vs time (Oct. 16, 2017).

The intensity vs time and zenith angle in J,H,Ks band is shown in Fig.18 which is drawn from data in Oct. 16, 2017. The data of Ks band was corrected. Seen from the Fig.20, the tendency of intensity vs zenith angle is same in three band of J,H,Ks and the value in zenith angle of 0 degrees is the minimum. The flux increases when the zenith angle becomes smaller.

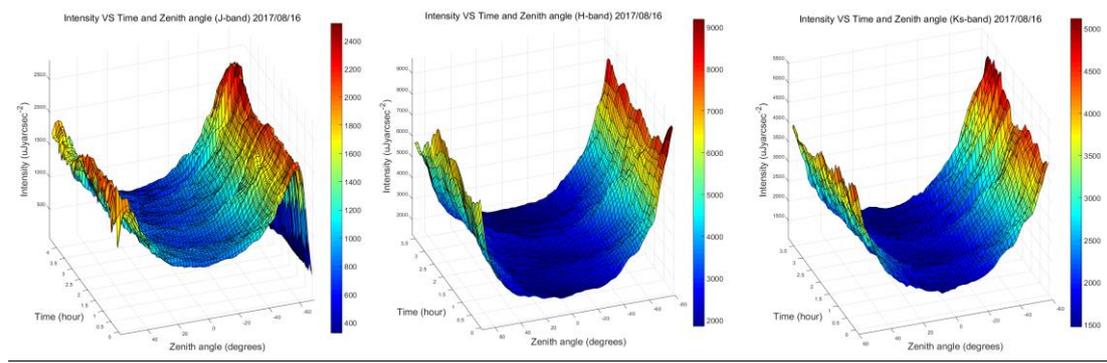

Fig.18. Intensity vs Time and Zenith angle in J,H,Ks band(Oct. 16, 2017)

The flux of zenith angle is shown in Table 8

Table 8. The flux of zenith angle of 0 degrees in J,H,Ks band

| flux($\mu Jy\ arcsec^{-2}$) | J-band | H-band | Ks-band |
|---|---|---|---|
| Mean | 942 | 2036 | 1084 |
| Max | 1136 | 2562 | 1317 |
| Min | 698 | 1610 | 842 |

# 6. Conclusion

A NIR sky brightness monitor is designed with InGaAs detectors which make the mechanics design simple for its low noise in J,H,Ks band without deep cooling using cyrocooler. The monitor consists of three instruments and each for one band which make the monitor more reliable. The blackbody calibration experiments in Lab was conducted to make the calibration easy done. The calibration results meet the theoretical analysis. The monitor was installed at Nagri Observatory with unattended operations. The first data of NIR sky brightness were obtained after installation since the end of July, 2017. The measurements show the monitor meets the demand of NIR sky brightness in sensitivity and resolution.

# Acknowledge

The authors greatly thank for discussion on optical system with Zheng-yang Li, Hai-ping Lu at National Astronomical Observatories/Nanjing Institute of Astronomical Optics and Technology, Chinese Academy of Sciences and Liang Chang at Yunnan Astronomical Observatories. This work was supported by the National Natural Science Funds of China under Grant No: 11603023, 11773026, the Fundamental Research Funds for the Central Universities (WK2360000003, WK2030040064), the Natural Science Funds of Anhui Province under Grant No: 1508085MA07, the SOC program (CHINARE2017-02-04), the Research Funds of the State Key Laboratory of Particle Detection and Electronics, the CAS Center for Excellence in Particle Physics.